\documentclass[]{interact}

\usepackage{epstopdf}      
\usepackage{natbib}        
\bibpunct[, ]{(}{)}{;}{a}{}{,}

\usepackage{booktabs}
\usepackage{amsmath}
\usepackage{graphicx}
\usepackage[hidelinks]{hyperref}

\providecommand{\tabnote}[1]{\par\vspace{2pt}\footnotesize #1}

\begin{document}

\articletype{ARTICLE}

\title{The Best Are Always the Best: COVID-19 Lockdown Stringency and
the Dispersion of Olympic Medal Outcomes}

\author{
\name{Fernando Delbianco\textsuperscript{a,b},
Federico Fioravanti\textsuperscript{c}\thanks{CONTACT Federico Fioravanti. Email: federico.fioravanti@univ-st-etienne.fr} and
Fernando Tohm\'e\textsuperscript{a,b}}
\affil{\textsuperscript{a}Departamento de Econom\'ia, Universidad Nacional
del Sur (UNS), Bah\'ia Blanca, Argentina;
\textsuperscript{b}Instituto de Matem\'atica Bah\'ia Blanca (INMABB),
CONICET, Bah\'ia Blanca, Argentina;
\textsuperscript{c}Universit\'e Jean Monnet Saint-\'Etienne, CNRS, Universit\'e Lyon 2, emlyon business school, GATE, 42023, Saint-\'Etienne, France.}
}

\maketitle

\begin{abstract}
We ask whether COVID-19 lockdown stringency altered national Olympic
performance between Rio 2016 and Tokyo 2020, using the Oxford Stringency
Index and the 99 countries that won a medal in either edition. As in
\citet{liu2024}, mean performance is unaffected: stringency is
insignificant in every OLS and ANOVA specification. The distribution is
not. Among the 84 non-traditionally dominant nations, medal changes are
three to six times more dispersed in high-stringency countries; the
difference is absent among dominant nations and concentrated in men's
events. A common shock left the competitive hierarchy intact while
sharply raising outcome uncertainty for smaller Olympic teams.
\end{abstract}

\begin{keywords}
COVID-19; Olympic Games; stringency index; outcome uncertainty;
medal dispersion
\end{keywords}

\vspace{4pt}
\noindent\textbf{JEL classification:} Z20; Z28; I18

\section{Introduction}

The Tokyo 2020 Games, postponed by a year and staged without spectators,
came after a months-long period of pandemics, which gave rise to an unusually heterogeneous episode of policy divergence. Between
Rio 2016 and Tokyo 2020, while some governments closed training facilities,
restricted movement and canceled domestic competition for months, others
barely intervened. Because those choices responded to epidemiological concerns and their domestic political response rather than to sport planning, they created the conditions for cross-country variation in the constraints faced by elite athletes over an entire Olympic cycle.

Two strands in the literature meet here. The first explains national Olympic success through population, income, host status, public policy and institutions \citep{hoffmann2002, bernard2004, johnson2004, mitchell2007, potts2014,
kufenko2021, rewilak2021}, and documents how persistent that success is once achieved \citep{csurilla2022}. 
The second uses the pandemic as a natural experiment in sport \citep{reade2021, destefanis2022,parnell2022,ehrlich2023,Delbianco2023}. 
The closest antecedent to the present work, \citet{liu2024}, combines them, showing that countries with higher COVID-19 mortality or stricter lockdowns won essentially the same number of medals predicted by their prior performance. The accurate predictions of the forecasting methodology in \citep{schlembach2022} also show that lockdowns had little effect on the results in the Olympic games.

Those previous results focused on the means of the distributions. We ask here a different question, closer to \citeauthor{baimbridge1998}'s \citeyearpar{baimbridge1998} treatment of
outcome uncertainty at the Olympics: did lockdown stringency change the
\emph{dispersion} of medal outcomes? The underlying economics is simple. A
common shock interacting with heterogeneous organizational capacity should
widen the outcome distribution without necessarily moving its center
\citep{williams2017}. Teams with deep organizational capital
\citep{debosscher2006} can be insulated by creating training bubbles, facility exemptions, and remote monitoring. The performance of Olympic programs without those resources would be closer to a lottery. The results should then reveal an asymmetry among the teams, with variance amplification concentrated among non-dominant nations.

We confirm this result and document exactly that asymmetry. Among
the 84 non-dominant countries in our sample, medal changes were three to
six times more dispersed under highly stringent lockdowns. Among the 15 historically
dominant nations, they were not. The effect is most notorious in men's events.

\section{Data and methods}

We draw our base data (medal counts by country, athlete sex, and event type) from the official IOC records for Rio 2016 and Tokyo 2020. The outcome is the change in the total medal count, $\textit{diffTotal} = \text{medals}_{2020} -
\text{medals}_{2016}$. The sample is the $N=99$ countries that won at
least one medal in either edition.

Policy exposure is obtained from the Oxford COVID-19 Government Response Tracker
stringency index \citep{hale2021}, a 0--100 composite of containment
measures, averaged over March 2020--July 2021, i.e.\ the preparation
window ending at the opening ceremony. A historical control,
$\textit{hist}_{2012}$ (total medals at London 2012), captures the 
strength and regression to the mean of national Olympic programs.

Fifteen countries are classified as historically dominant based on
cumulative medal count across the 1992--2012 Summer Games, corresponding
to all nations that won at least 15 medals at the 2012 London Games and
maintained top-tier standing across multiple preceding editions (USA,
Russia, UK, Germany, China, Australia, France, Japan, South Korea, Italy,
Cuba, Hungary, Netherlands, Canada, Spain). The remaining 84 constitute the
non-dominant subsample. Results are robust to a top-20 cut
($F[63,14]=0.268$, $p<.001$; $F[44,33]=0.278$, $p<.001$) but sensitive
to a top-10 cut (n.s.), where the Netherlands (diffTotal $= +17$,
low-stringency) enters the non-dominant pool and inflates low-stringency
variance. Table~\ref{tab:descriptives} reports the descriptive statistics.

For means, we estimate OLS of \textit{diffTotal} on the stringency index,
with and without $\textit{hist}_{2012}$, on both samples, and run one-way
ANOVA comparing high- and low-stringency groups at five thresholds: two
data-driven breakpoints obtained by ordering countries on the index and
applying the procedure of \citet{zeileis2002} (53.57 and 21.48), and the
sample median (49.07), mean (49.73) and first quartile (38.89).

For dispersion, we report the two-sample $F$ test of equality of variances,
$F = \text{var(low)}/\text{var(high)}$, so that $F<1$ signals greater
dispersion under high stringency. Because that test assumes normality
(Shapiro--Wilk: high-stringency group $W=0.947$, $p=.051$, borderline),
every result is confirmed with Brown--Forsythe tests (Levene's test
centered on the median, robust to non-normality); all key results hold
(Table~\ref{tab:variance}, note). Wilcoxon rank-sum tests find no
significant location differences at any threshold, consistent with a
pure variance effect. All computations are programmed in R.

Identification rests on stringency being set for epidemiological and
political reasons unrelated to Olympic planning. The main threat is that
richer countries locked down less and always invest more in elite sport. Adding
log GDP (2019, World Bank) and log GDP per capita as controls on the same
$N=99$ sample leaves the stringency coefficient insignificant
($\hat\beta = -0.003$, $p = .936$ with log GDP; $\hat\beta = 0.002$,
$p = .949$ with both), with VIF $<1.13$ throughout.

\begin{table}
\tbl{Descriptive statistics.}
{\begin{tabular}{@{}lrrrrr@{}}
\toprule
Variable & $N$ & $M$ & $SD$ & Min & Max \\
\midrule
\multicolumn{6}{@{}l}{\emph{Panel A: full sample ($N = 99$)}} \\[2pt]
Medal change (\textit{diffTotal})    & 99 & $ 1.19$ & $ 5.11$ & $-11$ & $ 18$ \\
Stringency index                     & 99 & $49.73$ & $16.84$ & $ 11$ & $ 87$ \\
2012 medals ($\textit{hist}_{2012}$) & 99 & $ 9.51$ & $17.74$ & $  0$ & $104$ \\[4pt]
\multicolumn{6}{@{}l}{\emph{Panel B: non-dominant subsample ($n = 84$)}} \\[2pt]
Medal change (\textit{diffTotal})    & 84 & $ 0.43$ & $ 3.30$ & $-11$ & $  9$ \\
Stringency index                     & 84 & $49.32$ & $17.23$ & $ 11$ & $ 87$ \\
2012 medals ($\textit{hist}_{2012}$) & 84 & $ 3.73$ & $ 4.27$ & $  0$ & $ 19$ \\
\bottomrule
\end{tabular}}
\label{tab:descriptives}
\tabnote{\emph{Note.} Medal change $=$ Tokyo 2020 $-$ Rio 2016. At the
mean stringency threshold (49.73): full sample, $n=48$ high and $n=51$
low; non-dominant subsample, $n=42$ in each group.}
\end{table}

\section{Results}

\subsection{No effect on mean performance}

Table~\ref{tab:ols} reports the OLS estimates. In the full sample the
stringency coefficient is $0.010$ ($SE = 0.031$, $p = .759$): ten
additional points of stringency (about six-tenths of a standard
deviation) move the expected medal change by one-tenth of a medal.
Adding $\textit{hist}_{2012}$ leaves it unchanged, and the non-dominant
subsample delivers a coefficient of the opposite sign and equally
insignificant. ANOVA at all five thresholds yields no significant mean
differences (all $p > .05$; the largest statistic is $F = 3.483$,
$p = .066$, at the upper breakpoint in the non-dominant subsample).

This replicates \citet{liu2024} on an independent design (Summer Games
only, a pre--post comparison around a single shock) and, as
\citet{abadie2020} argues, a precisely estimated null from a
theoretically motivated specification is informative in its own right:
the standard errors here rule out effects larger than roughly $0.07$
medals per stringency point.

\begin{table}
\tbl{OLS regressions predicting the change in Olympic medal count
(\textit{diffTotal}).}
{\begin{tabular}{@{}lcccc@{}}
\toprule
& \multicolumn{2}{c}{Full sample ($N = 99$)}
& \multicolumn{2}{c}{Non-dominant ($n = 84$)} \\
\cmidrule(lr){2-3} \cmidrule(lr){4-5}
Predictor & (1) & (2) & (3) & (4) \\
\midrule
StrIndex
  & $0.010$   & $0.007$     & $-0.007$  & $-0.007$ \\
  & $(0.031)$ & $(0.030)$   & $(0.021)$ & $(0.021)$ \\
  & $p=.759$  & $p=.818$    & $p=.750$  & $p=.735$ \\[4pt]
$\textit{hist}_{2012}$
  &           & $0.066^{*}$ &           & $-0.132$ \\
  &           & $(0.029)$   &           & $(0.085)$ \\
  &           & $p=.023$    &           & $p=.123$ \\[4pt]
Intercept
  & $0.720$   & $0.217$     & $0.762$   & $1.270$ \\
  & $(1.616)$ & $(1.596)$   & $(1.103)$ & $(1.141)$ \\
\midrule
$R^{2}$       & $.001$  & $.054$ & $.001$  & $.030$ \\
Adj.\ $R^{2}$ & $-.009$ & $.034$ & $-.011$ & $.006$ \\
\midrule
\multicolumn{5}{@{}l}{\emph{One-way ANOVA by stringency group (selected thresholds)}} \\[2pt]
Breakpoint ($>53.57$) & \multicolumn{2}{c}{$F=0.212$, $p=.646$}
                      & \multicolumn{2}{c}{$F=3.483$, $p=.066$} \\
Mean ($>49.73$)       & \multicolumn{2}{c}{$F=0.000$, $p=.993$}
                      & \multicolumn{2}{c}{$F=0.352$, $p=.555$} \\
\bottomrule
\end{tabular}}
\label{tab:ols}
\tabnote{\emph{Note.} Standard errors in parentheses. Models (1) and (3)
include StrIndex only; models (2) and (4) add $\textit{hist}_{2012}$.
$^{*}p<.05$.}
\end{table}

\subsection{Dispersion among non-dominant nations}

Table~\ref{tab:variance} turns to second moments. Among non-dominant
nations the variance ratio is $0.336$ at the upper breakpoint
($p<.001$), variance is three times larger under high stringency, 
$0.271$ at the mean threshold ($\approx 3.7\times$) and $0.164$ at the
first quartile ($\approx 6\times$). Levene and Wilcoxon rank-sum tests
agree at every significant threshold, and Shapiro--Wilk statistics
indicate approximate normality in both groups. The pattern replicates
when medal shares are used instead of counts.

\begin{figure}
\centering
\includegraphics[width=0.72\textwidth]{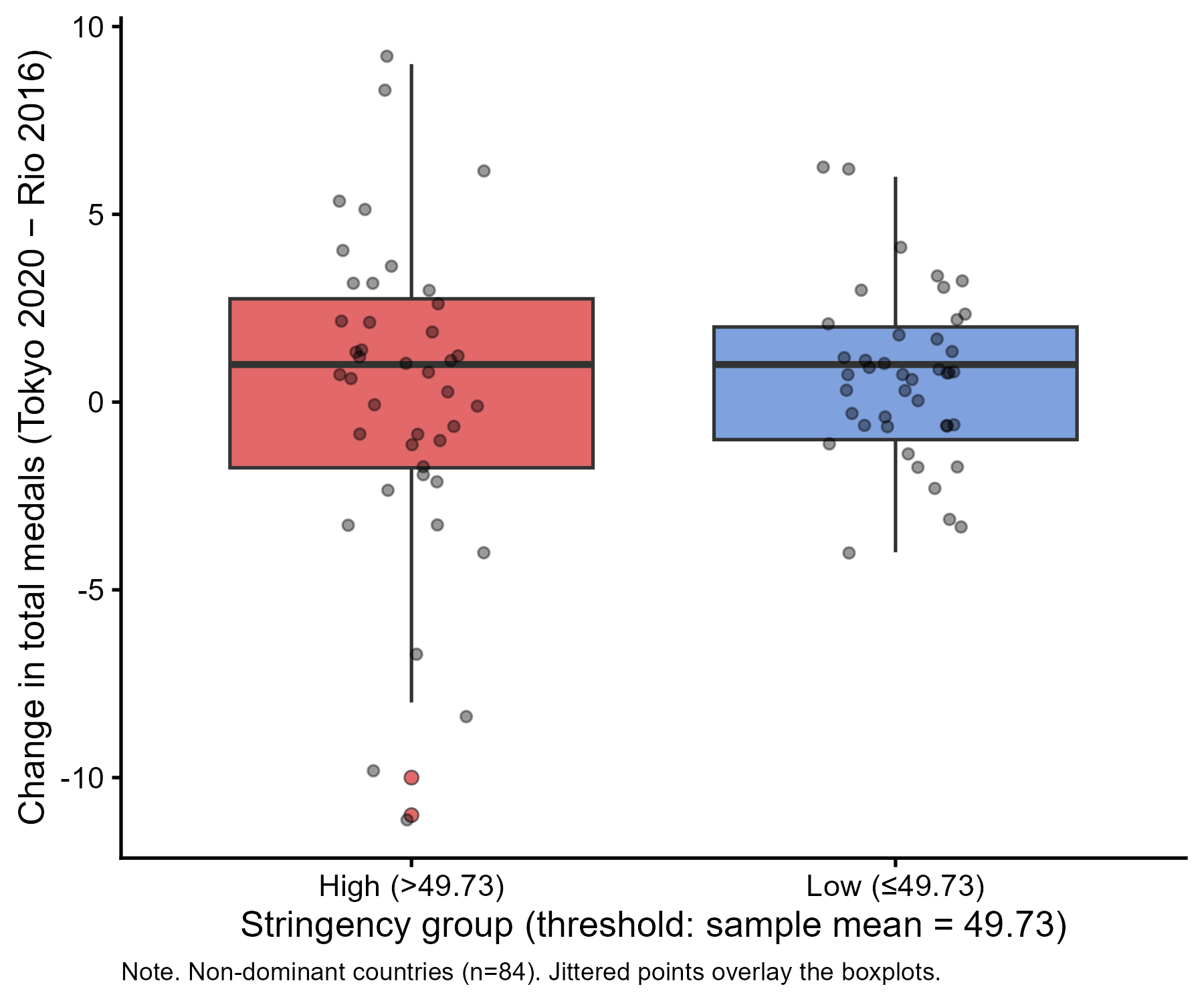}
\caption{Distribution of medal count changes by stringency group,
non-dominant countries ($n = 84$). Box plots compare \textit{diffTotal}
(Tokyo 2020 $-$ Rio 2016) above and below the mean stringency threshold
(49.73). The wider spread in the high-stringency group corresponds to the
variance ratio reported in Table~\ref{tab:variance}.}
\label{fig:boxplot}
\end{figure}

The predicted asymmetry holds. In the full sample, which adds the 15
dominant nations, the same tests are weaker and inconsistent (lower panel
of Table~\ref{tab:variance}): pooling insulated Olympic teams with exposed
ones attenuates the pattern, as implied by the organizational-capacity mechanism. Within the non-dominant subsample, the effect is significant for
individual events ($F[50,32] = 0.459$, $p = .013$) but not for team
events ($F[50,32] = 0.567$, $p = .070$), consistent with individual
athlete preparation being the operative channel.
Figure~\ref{fig:boxplot} displays the distributions.

\begin{table}
\tbl{$F$ tests for equality of variance in medal change by stringency group.}
{\begin{tabular}{@{}llrrrl@{}}
\toprule
Sample & Threshold & Cutoff & $F$ & $p$ & Interpretation \\
\midrule
Non-dominant & Breakpoint & 53.57 & 0.336 & $<.001$ & High stringency more dispersed \\
             & Mean       & 49.73 & 0.271 & $<.001$ & High stringency more dispersed \\
             & Q1         & 38.89 & 0.164 & $<.001$ & High stringency more dispersed \\
             & 2nd break  & 21.48 & 0.148 & $.074$  & Not significant \\
\midrule
Full sample  & Breakpoint & 53.57 & 0.579 & $.058$  & Marginal \\
             & Mean       & 49.73 & 0.726 & $.266$  & Not significant \\
             & Q1         & 38.89 & 0.371 & $.007$  & High stringency more dispersed \\
             & 2nd break  & 21.48 & 0.062 & $.015$  & High stringency more dispersed \\
\bottomrule
\end{tabular}}
\label{tab:variance}
\tabnote{\emph{Note.} $F = \text{var(low stringency)}/\text{var(high
stringency)}$; $F<1$ indicates greater dispersion under high stringency.
Degrees of freedom, non-dominant subsample: breakpoint $F(50,32)$, mean
$F(41,41)$, Q1 $F(22,60)$. Full sample: $F(59,38)$, $F(50,47)$,
$F(25,72)$. Shapiro--Wilk, non-dominant subsample: high stringency
$W=0.947$, $p=.051$; low stringency $W=0.969$, $p=.321$.
Brown--Forsythe tests (Levene centred on median), non-dominant subsample:
breakpoint $L(1,82)=7.14$, $p=.009$; mean $L(1,82)=6.61$, $p=.012$;
Q1 $L(1,82)=5.61$, $p=.020$. Full sample: breakpoint $L=2.53$, $p=.115$;
mean $L=1.29$, $p=.258$ (same pattern as $F$ tests).}
\end{table}

\subsection{Gender disaggregation}

Neither men's nor women's medal changes are predicted by stringency in any
specification (all $p > .40$). Women's outcomes are strongly path
dependent: London 2012 medals predict the change in women's medals in the
full sample ($\hat\beta = 0.102$, $SE = 0.016$, $p<.001$, $R^{2} = .30$),
with at most a marginal analogue for men. The dispersion effect is
confined to men's events (breakpoint $F[50,32] = 0.190$, $p<.001$, mean
threshold $F[41,41] = 0.250$, $p<.001$) and absent for women's
(breakpoint $F[50,32] = 1.166$, $p = .653$). Gender-asymmetric pandemic
effects are well documented in other domains \citep{alon2020}. Here the
asymmetry favors women, consistent with the institutionalized structures
that \citet{lowen2016} and \citet{berdahl2015} link to women's Olympic
success. Athlete-level evidence on pandemic distress
\citep{hakansson2020} suggests that individual-level mechanisms may also play a role, and are
a matter of future work.

\section{Conclusion}

Lockdown stringency did not shift the center of the Olympic medal
distribution, but it stretched the tails for countries outside the
traditional elite. The first result confirms \citet{liu2024}: the
competitive hierarchy survived. The second qualifies it. Reading the null hypothesis
as ``the pandemic did not matter'' conflates a mean with a distribution.
For an Olympic team outside the traditional elite, a three- to six-fold
increase in the variance of medal outcomes is a large change in risk even
when the expected value is unchanged, and precisely the change that
mean-based analyses, including forecasting models
\citep{schlembach2022}, are not designed to detect.

Read through the outcome-uncertainty tradition that
\citet{baimbridge1998} brought to the Olympics, COVID-19 acted as an
uncertainty shock. Greater uncertainty raises the option
value of participation for marginal national teams while making returns to
investment in elite sport less predictable, an ambiguity already detected in the analysis of the relation between institutions and inequality in Olympic outcomes
\citep{kufenko2021}.


\section*{Disclosure statement}

No potential conflict of interest was reported by the authors.

\section*{Funding}

This work was supported by the French National Research Agency under Grant ANR-24-EXMA-0001 PEPR MathsVivES CONDORCET.

\section*{Data availability statement}

Medal data are publicly available from the International Olympic
Committee. The Oxford COVID-19 Government Response Tracker data are
available at \url{https://github.com/OxCGRT/covid-policy-tracker}.
Replication code is available from the corresponding author on request.

\bibliographystyle{tfcad}
\bibliography{references}

\end{document}